\documentclass[twocolumn,twoside,aps,pra,showpacs,superscriptaddress,floatfix]{revtex4}
\usepackage{epsf,epsfig,amsfonts,amsmath,amssymb,graphicx,times}
\usepackage[utf8]{inputenc}
\usepackage[OT1]{fontenc}
\usepackage{mathtools,color,stmaryrd,hyperref}

\begin{document}
\title{Unidimensional continuous-variable quantum key distribution}
\author{Vladyslav C. Usenko}
\email{usenko@optics.upol.cz}
\affiliation{Department of Optics, Palacký University, 17. listopadu 50,  
  772~07 Olomouc, Czech Republic}
\affiliation{Bogolyubov Institute for Theoretical Physics of National Academy of Sciences,
Metrolohichna st. 14-b, 03680, Kiev, Ukraine}
\author{Frédéric Grosshans}
\email{frederic.grosshans@u-psud.fr}
\affiliation{Laboratoire de Photonique Quantique et Moléculaire, CNRS, ENS Cachan, Université Paris Saclay, F-94235 Cachan, France}
\affiliation{Laboratoire Aimé Cotton, CNRS, Univ. Paris-Sud, ENS Cachan,
    Universite Paris Saclay, F-91405 Orsay, France}
\begin{abstract}
We propose the continuous-variable quantum key distribution protocol based on the Gaussian modulation of a single quadrature of the coherent states of light, which is aimed to provide simplified implementation compared to the symmetrically modulated Gaussian coherent-state protocols. The protocol waives the necessity in one of quadrature modulations and the corresponding channel transmittance estimation. The security of the protocol against  collective attacks in a generally phase-sensitive Gaussian channels is analyzed 
and is shown achievable upon certain conditions. Robustness of the protocol to channel imperfections is compared to that of the symmetrical coherent-state protocol. The simplified unidimensional protocol is shown possible at a 
reasonable quantitative cost in terms of key rate and of tolerable channel excess noise.
\end{abstract}
\maketitle

\section{Introduction}

Over the last three decades, quantum key distribution (QKD) \cite{qkd} has emerged as 
a way to ensure the security of a secret key through the very nature of quantum states
distributed between trusted parties. Recent developments in this field are concerned with the continuous-variables (CV)  coding of key bits, \cite{CVreview, Ralph99, Hillery, sqGP,sqCLvA,sqFurrer+,sqFurrer, Madsen12, sqXP, GG02, GG02b, cohLeuchs, cohXP1, noswitch, noswitchexp, cohXPlast, cohXPnext, binary1D} in particular, 
the Gaussian modulation of the field quadratures of 
squeezed \cite{sqGP,sqCLvA,sqFurrer+,sqFurrer, Madsen12, sqXP}
 and coherent states of light \cite{GG02, GG02b, cohLeuchs, noswitch, noswitchexp, cohXP1, cohXPlast, cohXPnext}.
Coherent state protocols are more promising experimentally 
\cite{cohXP1, cohXPlast, cohXPnext}, 
and the main goal of the present paper is to propose their further simplification. 
In particular, all published coherent-state protocols suppose a symmetrical amplitude 
and phase quadrature modulation 
[with the exception of the binary Zhao-Heid-Rigas-L\"utkenhaus 2009 (ZHRL09) protocol \cite{binary1D}
However, an asymmetric protocol allows Alice to use one modulator (e.g. an amplitude modulator) instead of two. This would
reduce the complexity and the cost of Alice’s apparatus. Furthermore, the amplitude modulator used in a symmetric CV QKD apparatus needs to have a strong extinction ratio, in order to avoid creating a "hole” in the center of the Gaussian probability distribution \cite{discussion}. On the other hand, a simple single-quadrature amplitude modulation does not have this need, and the use of more standard (and cheaper) modulators becomes possible.

Thus, in the present paper we propose the unidimensional (UD) CV QKD protocol based on the Gaussian single-quadrature 
modulation of coherent states of light. 
We show the security of the protocol in a general phase-sensitive channel
restricting eavesdropper only by the physicality constraints and keeping to the pessimistic worst-case assumptions. 
Then we compare the UD protocol to the standard coherent-state protocol and discuss the possible extensions. 
Our paper thus continues the tendency of technical simplification of the QKD protocols which was started in \cite{cheapBB84}, where the low cost and compact discrete variable QKD system was proposed.

In Sec. \ref{sec:protocol}, we present the UD CV QKD protocol, 
and we analyze its security in Sec. \ref{sec:security}.
We then look at its performance in the common case of symmetric quantum channels and
compare it to the usual symmetric protocol in Sec. \ref{sec:performance}.

\section{Unidimensional protocol}
\label{sec:protocol}

\begin{figure}
\includegraphics[width=0.9\columnwidth]{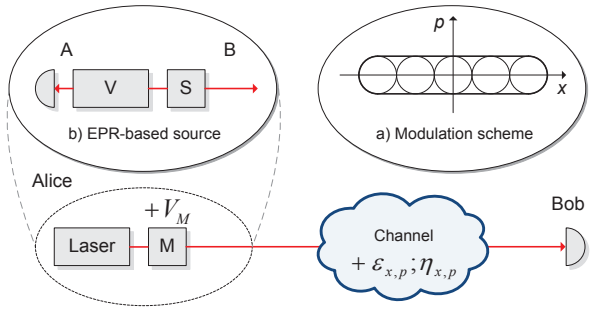}
\caption{(Color online) Scheme of the UD coherent-state protocol. Alice prepares a coherent state using a laser source and then modulates the state by
displacing it along the modulated quadrature using modulator M so that the modulation variance is $V_M$. The states travel through an untrusted generally phase-sensitive channel to a remote party Bob, who performs homodyne measurement of the modulated quadrature. (a) Mixture of modulated coherent states on a phase space 
(assuming $x$ quadrature was modulated). (b) Equivalent entanglement-based scheme using a two-mode squeezed vacuum source, mode A is measured by Alice using a homodyne detector, mode B is squeezed on the squeezer S and sent to the channel.
\label{scheme}}
\end{figure}

The central idea of the protocol is to modulate a single quadrature  of coherent states, 
in contrast to the usual coherent-state protocols, where two quadratures are simultaneously modulated. 
This should provide simplified implementation, at the price of slightly degraded performances,
as we show below.
The scheme of the protocol is given in Fig. \ref{scheme}. 
One of the trusted sides, Alice, 
produces coherent states, \emph{e.g.} with a laser source.
Then she applies modulation in one of the quadratures (denoted as x), using modulator M, 
and displaces each coherent state according to a random Gaussian variable with 
displacement variance $V_M$. 
With no loss of generality we further assume the modulated quadrature $x$ to be the 
amplitude quadrature. In this case the displacement can be performed by an intensity 
modulator.
The mixture of the modulated states thus forms a ``sausage'' on a phase-space 
[see Fig. \ref{scheme} (a)]. 
Its thickness is the quadrature variance of a coherent state, \emph{i.e.} 1 shot 
noise unit (SNU), and its length is $\sqrt{V_M+1}$ SNU.
The states are then sent to the remote trusted party Bob through a generally 
phase-sensitive channel with transmittance $\eta_x$, $\eta_p$, and excess noise 
$\epsilon_x$, $\epsilon_p$ in $x$, and $p$ quadrature, respectively. 
Bob performs a homodyne measurement of the modulated quadrature, using a homodyne detector, measuring most of the time the $x$quadrature, and sometimes 
measuring the $p$quadrature. 
This basis switching should be performed often enough to gather statistics on the 
properties of the channel in the $p$-quadrature. 
However, in the asymptotic limit of many repetitions studied here, these measurements 
can be a vanishing fraction of the total data set and have a negligible impact on the key rate
\cite{EffQKD}.
After a sufficient number of runs, Alice and Bob analyze the security and extract a
secret key from the $x$-quadrature data 
using a reverse-reconciliation procedure \cite{GG02b, cohXP1}.

In the following section we estimate the security region of the UD protocol and compare it to the standard coherent-state based protocol.

\section{Security of the protocol}
\label{sec:security}
Let us study the protocol in detail and estimate its applicability. 
\subsection{Computing the Key Rate from the Covariance Matrix}
The study of security of CV QKD protocols including finite size effects \cite{finitesize, finitesize2} 
and coherent attacks is an ongoing research 
program\cite{sqFurrer+, sqFurrer, coherentatt, coherentatt2, coherentatt3}. 
Very recently \cite{composable}, Leverrier has shown for the first time the composable security
against general attacks for a CV QKD coherent-state protocol.
It was shown that for the symmetrically modulated coherent-state protocol, 
the optimal attacks are the Gaussian attacks and the corresponding secret 
key rate tends to the one obtained for
Gaussian collective attacks \cite{collective1, collective2, collective3} 
for high number of pulses.

We will compute the asymptotic key rate of our protocol against collective attacks.
An approach similar to \cite{composable} can likely be applied to extend this security
to a general attack, but this work is kept for future research. 

The extremality of Gaussian states \cite{extremality} and subsequent optimality of Gaussian attacks \cite{collective1, collective2} allows one to use the powerful covariance matrix formalism 
to estimate the amounts of information leaking to a potential eavesdropper under given channel conditions.

In the case of collective attacks the lower bound on the key rate is given by the difference between classical (Shannon) mutual 
information, available to the trusted parties (A and B), and the upper bound on the information extractable from the state possessed by an eavesdropper (E) and conditioned by the measurement results of the reference side of the classical post processing algorithms, i.e., in the case of reverse reconciliation the lower bound reads:
\begin{equation}
\label{kr}
K=I_{AB}-\chi_{BE},
\end{equation}
where $\chi_{BE}=S(E)-S(E|x_B)$ 
is the Holevo quantity \cite{Holevo_bound}, being the capacity of a bosonic channel between an eavesdropper (E) and the reference side of the information reconciliation (B), quantified as the difference of von Neumann entropy $S(E)$ of the state, available to an eavesdropper, and the entropy $S(E|x_B)$ of the eavesdropper state, conditioned by the measurement results of the remote trusted party B \cite{collective1,collective2}. The positivity of the lower bound (\ref{kr}) means that the post-processing algorithms are able to distill the secure key 
\cite{cktheorem, devtak}, 
i.e., that the protocol is secure under given channel conditions. 
In the cases where channel noise is present, 
the collective attack can be accessed through the assumption that
the eavesdropper holds the purification of the state, shared between A and B, thus the entropies of the sub states of the generally pure state are equal: $S(E)=S(AB)$ and $S(E|x_B)=S(A|x_B)$. The calculation of the von Neumann entropies, contributing to the Holevo quantity, is done, using the covariance matrix formalism, explicitly describing the Gaussian states, through the symplectic eigenvalues 
$\lambda_{1,2}$ and $\lambda_{\text{cond}}$ of the respective covariance matrices $\gamma_{AB}$ prior to and $\gamma_{A|x_B}$ after the measurement so that
\begin{equation}\label{holevo1}
\chi_{BE}=G\left(\frac{\lambda_1-1}{2}\right)+G\left(\frac{\lambda_2-1}{2}\right)
              -G\left(\frac{\lambda_{\text{cond}}-1}{2}\right),
\end{equation}
where $G(x)=(x+1)\log (x+1)-x\log x$ \cite{footnote1} 
is the bosonic entropic function \cite{negexpr}. 

\subsection{Which Covariance Matrices are Physical ?}
To analyze the security of the protocol we switch to the equivalent entanglement-based (EPR) scheme \cite{equiv}, which allows the explicit description of trusted modes and their correlations. For the UD protocol such a scheme can be built, by taking a two-mode squeezed vacuum state of variance $V$ and squeezing one of its modes with the squeezing parameter $-\log{\sqrt{V}}$, resulting in the covariance matrix:
\begin{equation}
\label{inputstate}
\gamma_{AB}=\begin{bmatrix}
	V & 0 & \sqrt{V(V^2-1)} & 0 \\
	0 & V & 0 & \mathllap{-}\sqrt{\frac{V^2-1}{V}} \\
	\sqrt{V(V^2-1)} & 0 & V^2 & 0 \\
	0 & \mathllap{-}\sqrt{\frac{V^2-1}{V}} & 0 & 1
\end{bmatrix}
\end{equation}
As stated above, the modulated quadrature is the intensity quadrature $x$.
If Alice performs a homodyne measurement 
on the mode A, then the coherent state is conditionally prepared and is effectively 
sent to the remote party Bob. The EPR-scheme is then equivalent to the Gaussian displacement of coherent states along the $x$ quadrature with the variance $V_M=V^2-1$. As the states travel through the noisy and lossy channel, the 
covariance matrix is transformed according to the channel parameters. 
However, since there is no modulation in the $p$ quadrature, 
the correlation, and, respectively, the channel transmittance in $p$ cannot be estimated. 
The remote party can therefore only  measure 
 the variance of the channel output in $p$. Thus, 
 generally, the covariance matrix after the channel in terms of the modulation variance $V_M$ has the form:
\begin{multline}
\label{outputstate}
\gamma_{AB}'=\\
 \begin{bmatrix}
    \sqrt{1+V_M} & 0 & \sqrt{\eta_xV_M}(1+V_M)\mathrlap{^{\frac{1}{4}}} & 0 \\
    0 & {\sqrt{1+V_M}}  & 0 & C_p \\
    \sqrt{\eta V_M}(1+V_M\mathrlap{)^{\frac{1}{4}}} & 0 & 1+\eta_x(V_M+\epsilon_x) & 0 \\
    0 & C_p & 0 & V_p^B
 \end{bmatrix}
\end{multline}
where $\eta_x$ and $\epsilon_x$ are, respectively, the channel transmittance and 
excess noise, estimated by the trusted parties through the measurement of the $x$ quadrature; 
$V_p^B$ is the output variance of the mode B in the $p$ quadrature, 
which is measured at the remote side, and
$C_p$ is the correlation between trusted modes in the $p$ quadrature, being unknown due to the fact that the quadrature is not modulated, which means that the channel transmittance is not estimated in $p$. 

The covariance matrix of the state, conditioned by Bob's measurement in $x$ is given by
\begin{equation}
\gamma_{A|x_B}=\gamma_A-\sigma_{AB}(X \gamma_B X)^{\mathit{MP}}\sigma_{AB}^T,
\end{equation}
where $\gamma_A$, $\gamma_B$ are the submatrices of the covariance matrix 
$\gamma'_{AB}$ (\ref{outputstate}), describing the modes A and B individually; $\sigma_{AB}$ is the submatrix of (\ref{outputstate}), which characterizes correlation between modes A and B; $\mathit{MP}$ stands for Moore Penrose (pseudo-) inverse of a matrix, and
\begin{equation}
X =
\left( \begin{array}{cc}
1 & 0 \\
0 & 0
\end{array} \right).
\end{equation}
In the general case the conditional matrix is thus given by
\begin{equation}
\label{condmat}
\gamma_{A|x_B} =
 \begin{bmatrix}
 \frac{\sqrt{V_M+1}(1+\eta_x\epsilon_x)}{1+\eta_x(V_M+\epsilon_x)} & 0 \\
 0 & \sqrt{1+V_M} 
\end{bmatrix},
\end{equation} 
Now let us estimate the lower bound on the key rate (\ref{kr}) for our single quadrature protocol. Shannon mutual information between the trusted parties is easily calculated from the first diagonal elements of the matrices $\gamma_A$ and $\gamma_{A|x_B}$:
\begin{equation}
\label{mutinfgen}
I_{AB}=\frac{1}{2}\log{\frac{V_A}{V_{A|B}}}=\frac{1}{2}\log{\bigg(1+\frac{\eta_x V_M}{1+\eta_x\epsilon_x}\bigg)}
\end{equation}

On the other hand, the estimation of the Holevo quantity $\chi_{BE}$, representing the upper bound on information, available to an eavesdropper, should be done from the whole state and, thus, depends on the unknown 
correlation parameter $C_p$. 
However, this unknown parameter is bounded by the requirement of the physicality of the state, which is given by the Heisenberg uncertainty principle, in terms of the covariance matrices being \cite{negexpr}
\begin{equation}
\label{uncert}
\gamma_{AB}'+i\Omega \geq 0,
\end{equation}
where $\Omega$ is the symplectic form
\begin{equation}
\Omega=\bigoplus_{i=1}^{n}\omega\, , \quad \omega=
\left(\begin{array}{cc}
0&1\\
-1&0
\end{array}\right)\, . \label{symform}
\end{equation}
This equation imposes physical constraints on the possible values of $C_p$.
Such constraint in the general case of noise present in both quadratures is given by the parabolic equation on the $\{V_p^B,C_p\}$ plane:
\begin{equation}
\label{cpmax}
(C_p-C_0)^2\le\frac{V_M}{(1+V_M)^{\frac{1}{2}}}(1-\eta_xV_0^B)(V_p^B-V_0^B)
\end{equation}
with vertex $(V_0^B,C_0)$, defined as:
\begin{equation}
\label{vp0}
V_0^B=\frac{1}{1+\eta_x\epsilon_x}
\end{equation}
and
\begin{equation}
\label{c0}
C_0=-\frac{V_0^B\sqrt{\eta_xV_M}}{(1+V_M)^{\frac{1}{4}}}.
\end{equation}

The first part of the Holevo quantity, $S(AB)$, can be calculated from the symplectic eigenvalues $\lambda_{1,2}$ that are given by the square roots of the solutions of equation
\begin{equation}
z^2-\Delta z + \det\gamma_{AB}'=0,
\end{equation}
where $\Delta=\det\gamma_A+\det\gamma_B+2\det\sigma_{AB}$ is the second 
symplectic invariant, the first one being $\det\gamma_{AB}'$. The second part, $S(A|x_B)$, is calculated from 
$\lambda_{\text{cond}}=\sqrt{\det\gamma_{A|x_B}}$. 
This allows one to analytically derive the lower bound on the key rate and find the
security bounds in terms of the unknown correlation $C_p$ upon given (measured) $V_p^B$. 

\begin{figure}[tb]
\includegraphics[width=0.9\columnwidth]{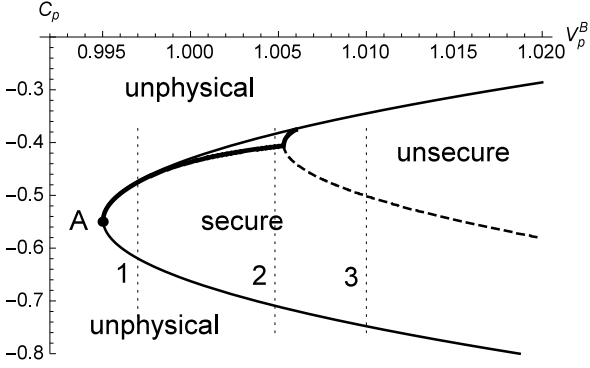}
\caption{Physicality (solid line) and security within the physicality (dashed line) regions of the UD protocol. The pessimistic value of $C_p$, which minimizes the key rate, is given as a bold solid line. Modulation variance $V_M=10$, channel transmittance in x, $\eta_x=0.1$, noise in x, $\epsilon_x=5\%$ SNR. Point $A=(C_0,V_0^B)$ denotes the vertex of the parabola, described by (\ref{cpmax}). The lines 1, 2, and 3 correspond to the key rate dependencies given in Fig. \ref{KRvsCP}. 
\label{cbounds}}
\end{figure}

The corresponding physicality region and security within physicality in terms of 
the correlation $C_p$ are given in Fig. \ref{cbounds}.

\begin{figure}[tb]
\includegraphics[width=0.9\columnwidth]{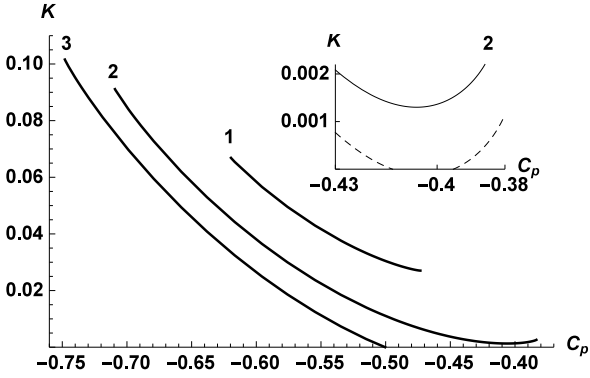}
\caption{Key rate secure against collective attacks versus correlation $C_p$ for different values of variance $V_p^B$, corresponding to the lines 1, 2, and 3 in Fig. \ref{cbounds}. The rest of the channel parameters are the same as in the mentioned figure. Inset demonstrates the dependence of line 2 in the smaller scale. For comparison the line corresponding to $V_p^B=1.00535$ is given as dashed, demonstrating the particular case when security is lost and then restored.
\label{KRvsCP}}
\end{figure}

It is evident from the graph, that there exists a region of $V_p^B$, when the protocol is secure for any $C_p$. In this region the UD protocol can be indeed implemented without the necessity of the correlation estimation in $p$, since 
no physically valid collective attack can break the security. 
For higher values of $V_p^B$ the protocol cannot be implemented, 
since it would only be secure for some values of $C_p$, but Alice and Bob cannot 
estimate the latter quantity. 
Such a behavior can be clearly observed at the graphs in Fig. \ref{KRvsCP}, where dependence of the lower bound on the key rate on the correlation $C_p$ is given for different values of $V_p^B$, corresponding to the respective lines in Fig. \ref{cbounds}.

When the channel excess noise $\epsilon_x>0$ increases, the physicality region of the state after the channel expands, which allows Eve to perform stronger attacks.

\subsection{Worst-Case $C_p$ and Key Rate}
Counter-intuitively, the key rate is not always a monotonously decreasing function of the
correlation $|C_p|$.
Indeed, it can be seen from Fig. \ref{KRvsCP} that upon certain values of variance $V_p^B$ the lower bound on the key rate can have a local minimum within the security region. Moreover, the security can be even lost and restored (see the dashed line at the inset in Fig. \ref{KRvsCP}). 

However, when the channel excess noise added in $p$-quadrature is small 
(i.e., when $V_p^B$ is close to 1), the key rate is a monotonously decreasing function of the correlation $|C_p|$ 
(as can be also seen in Fig. \ref{KRvsCP}) 
in most of the physicality region, and the pessimistic value
for $C_p$ is typically the highest physically valid negative value $C_p^{\max}$, 
which saturates inequality (\ref{cpmax}). 

As the noise increases, the pessimistic value of $C_{p}$ gets lower than $C_p^{\max}$ and must be found numerically. We thus consider the security region of our protocol as laying along the pessimistic value of $C_p$ (given as a bold line in Fig. \ref{cbounds}) from $C_0$ to $V_p^{B,\max}$, where physicality and security regions cross.
In this case, a key rate computed at $C_p^{\max}$ is greater than the lower bound on the real key
rate and is therefore too optimistic. However, when the pessimistic value of $C_p$ is inside the parabola,
the $\partial K/\partial C_p=0$ at this point and the pessimistic value is usually close to  $C_p^{\max}$. 
This explains why this upper bound, computed below, is often a good approximation.

The parabola bounding the physicality region corresponds to a state saturating the
Heisenberg inequality \eqref{uncert}. Therefore, one of the symplectic eigenvalues
$\lambda_2=1$ and $\lambda_1=\sqrt{\det\gamma_{AB}'}$ and Eq. \eqref{holevo1} 
becomes
\begin{equation}
  \chi_{BE}\left(\tfrac12\sqrt{\det\gamma_{AB}'}-\tfrac12\right)
           -G\left(\tfrac12\sqrt{\det\gamma_{A|x_B}}-\tfrac12\right)
\end{equation}
when $C_p=C_p^{\max}$.

When $V_M\gg1$, i.e., in the strong-modulation limit, $\det\gamma_{AB}'\gg1$ 
and one can use the expansion of the bosonic function 
$G\left(\tfrac12(\lambda-1)\right)= \log\lambda+\log\tfrac e2
 -\tfrac{\log e}{6\lambda^2}+O\left(\tfrac1{\lambda^4}\right)$,
to derive the following expression for the key rate upper bound:
\begin{align}
 K_{V_M\rightarrow\infty}
  &\begin{multlined}[t]
   \lesssim\tfrac12\log\tfrac{\eta_x}{1-2\eta_x+\eta_xV_p^B+\eta_x\epsilon_x+2\sqrt{D}}
     -\log\tfrac e2\\+G\left(\tfrac12(\sqrt{\tfrac1{\eta_x}+\epsilon_x}-1)\right)
     +O(\tfrac1{\sqrt{\eta_x}V_M})
   \end{multlined}\\
 \text{with }D&=\eta_x(1+\eta_x\epsilon_x-\eta_x)(V_p^B(1+\eta_x\epsilon_x)-1),
\end{align} 
where $\lesssim$ can be replaced by $\simeq$ when $C_p^{\max}$ is indeed the worst $C_p$.
If, furthermore, we are in the strong loss limit, where $\eta_x\ll1$ \cite{footnote2}
  and $V_p^B$ is close to 1, 
one can expand the remaining bosonic function and obtain 
\begin{equation}
 K_{\substack{V_M\rightarrow\infty\\\eta_x\ll1}}
   \lesssim \left[\left(\tfrac13+\tfrac{1-V_p^B}2\right)\eta_x - \sqrt D\right]\log e
   +O(\eta^2+\tfrac1{\sqrt{\eta_x}V_M})
   \label{KVmasym}
\end{equation}

In the following Section we analyze the security of the UD protocol in the typical phase-insensitive Gaussian channels.

\section{Performance for Symmetric Quantum Channels}
\label{sec:performance}

In  typical communication channels, 
one expects values of loss and excess noise in both quadratures to be symmetric. 
 In this regime,  $\eta_x=\eta_p\equiv\eta$, 
$\epsilon_x=\epsilon_p\equiv\epsilon$, and therefore, $V_p^B=1+\eta\epsilon$.
The previous equations then become
\begin{gather}
 \label{approx_sym0}
 K_{V_M\rightarrow\infty}^{\text{sym}}
  \begin{multlined}[t]
   \lesssim\tfrac12\log\tfrac{\eta}{1-\eta+\eta\epsilon+\eta^2\epsilon+2\sqrt{D}}
     -\log\tfrac e2\\+G\left(\tfrac12(\sqrt{\tfrac1{\eta}+\epsilon}-1)\right)
      +O(\tfrac1{\sqrt{\eta_x}V_M})
   \end{multlined}\\
 \text{with }D=2\eta^2\epsilon(1+\eta\epsilon-\eta)(1+\tfrac12\eta\epsilon).\\
   K_{\substack{V_M\rightarrow\infty\\\eta\ll1}}^{\text{sym}}
     \lesssim  (\tfrac13 - \sqrt {2\epsilon})\eta\log e +O(\eta^2+\tfrac1{\sqrt{\eta_x}V_M})
		\label{approx_sym}
\end{gather}

Note that Eqs. (\ref{approx_sym0})-(\ref{approx_sym}) describe well the lower bound on the key rate if the losses or noise in the channel are low, i.e., $\eta \to 1$ or $\epsilon \to 0$, otherwise they give the result exceeding the lower bound on the key rate, and the latter needs to be calculated numerically using the pessimistic $C_p$ within the physicality region.

We now compare the UD CV QKD protocol 
with the standard symmetrical modulation protocol GG02 \cite{GG02, GG02b, collective1,
 collective2}
used over the same channel.
We first assume a noiseless lossy channel, where $\epsilon=0$. 

In this case, Eq. \eqref{cpmax} becomes $C_p=C_0$ and Eq. \eqref{KVmasym} gives therefore 
the key rate for our protocol. 
It becomes,
for $V_M\to\infty$ 
\begin{equation}
 K_{\substack{V_M\rightarrow\infty\\\epsilon=0}}^{\text{sym}}
  =\frac1{2\sqrt\eta}\log\frac{1+\sqrt\eta}{1-\sqrt\eta}-\log e
   +O(\tfrac1{\sqrt{\eta_x}V_M})
\end{equation}
Its low transmission limit rate is $\frac\eta3\log e$, slightly smaller than
the key rate
 of the standard coherent-state protocol in the high modulation limit, 
 given by \cite{grosshans2005}
\begin{equation}
K_{V_M\to\infty}^{\text{GG02}}=-\tfrac12\log{(1-\eta)}
  \simeq\frac\eta2\log e \text{ for }\eta\ll1.
\end{equation}

\begin{figure}
\includegraphics[width=0.9\columnwidth]{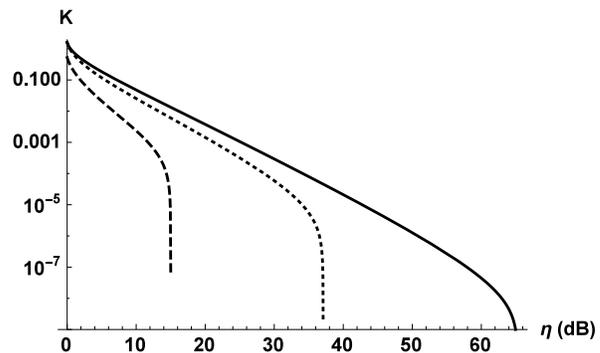}
\caption{Typical dependence of the key rate on loss (in dB scale) upon symmetric
channel excess noise $\epsilon=5\%$ SNU. (Solid line) Symmetrical coherent-state
protocol; (dashed line) UD protocol with correlation
estimation in $p$; (dotted line) UD protocol without correlation
estimation in $p$. Modulation variance $V_M=100$.
\label{comparison1}}
\end{figure}

\begin{figure}
\includegraphics[width=0.9\columnwidth]{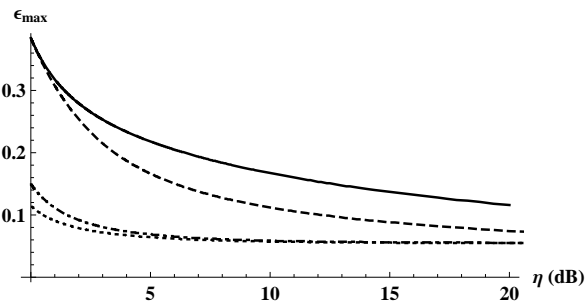}
\caption{Typical profile of the security region in terms of maximal tolerable channel excess noise $\epsilon$ versus channel loss (in dB scale). (Solid line) Symmetrical coherent-state protocol; (dashed line) UD protocol with channel estimation in $p$; dotted line: UD protocol without channel estimation in $p$; (dot-dashed line) optimistic evaluation of UD protocol without channel estimation in $p$ assuming $C_p=C_p^{max}$. Modulation variance $V_M=100$.
\label{comparison2}}
\end{figure}

In the general case, however, the channel noise is present and reduces
 the security of the protocol. The results of the calculations in this case are given in Fig. \ref{comparison1} in terms of the lower bound on the key rate upon fixed channel excess noise and in Fig. \ref{comparison2} in terms of the maximum tolerable channel excess noise versus channel loss upon strong modulation $V_M=100$. Evidently, the UD protocol demonstrates higher sensitivity to channel excess noise, which is the cost of technical simplification, but still provides the reasonable security region in terms of channel excess noise, even in the pessimistic assumption of the strongest physically possible collective attack. We also provide comparison with the case when the worst-case $C_p$ is not estimated numerically but is optimistically taken as a bound to physicality $C_p^{max}$ (so that the key rate is approximately given by (\ref{approx_sym}) in the limit of strong modulation), the respective curve is given as the dot-dashed line in Fig. \ref{comparison2}. It is evident from the plot that the optimistic assumption of the physicality-bounded $C_p^{max}$ gives the same security bounds as the pessimistic one when the channel attenuation goes below few dB. In this regime it is sufficient to bound security by the physicality condition.

Zhao \emph{et al.} have also introduced a 
single-modulation protocol, ZHRL09. Contrary to our independently developed protocol, it uses a binary modulation,
simplifying even more the protocol implementation. 
However, its sensitivity to excess noise is orders of magnitude below the tolerable excess
noise of the protocol presented here: 
An excess noise as small as $\epsilon=3\times10^{-3}$ does not allow any positive key rate
beyond 1 dB losses, and $\epsilon\simeq10^{-3}$ does not allow one to go beyond 4 dB losses. 
This extreme sensitivity renders ZHRL09 useless in practice. 
In their conclusion, Zhao \emph{et al.} attribute this sensitivity to the binary
modulation and predict that a Gaussian modulation would solve this problem.
The present paper indeed proves this conjecture.

For the sake of comparison we also analyzed the protocol, in which no information is extracted from the $p$ quadrature, but some modulation and measurement is performed to estimate the channel transmittance (and, equivalently, the correlation) in $p$. This intermediate protocol provides the security region, which lays in between the symmetrical and completely asymmetrical counterparts, but requires 
modulation in both quadratures.
Its main interest it theoretical, since it allows one to split the origin of the 
performance degradation of our protocol compared to GG02 between the degradation due
to the asymmetric modulation and the one
due to  incomplete channel estimation.

Another possible option to improve the UD protocol could be the noise addition in $p$ to decouple the eavesdropper from the remote trusted party. However, it widens the physicality region, allowing for the stronger collective attacks, and thus, if the noise is strong enough, security is always broken before the physicality bound, meaning that additional noise in $p$ makes the protocol inapplicable. Additionally, if the channel estimation in $p$ is performed, then the protocol shows the same performance as the standard squeezed-based protocol \cite{sqCLvA,GG02b,collective1, collective2}, 
since the homodyne detection on A projects the two-mode state on the single-mode squeezed state, getting more squeezed as the noise in $p$ increases.

Further analysis of the protocol will include consideration of reduced post-processing efficiency \cite{reconciliation}, composable security \cite{composable},
and finite-size effects \cite{finitesize,finitesize2}, which, however, depend on the signal-to-noise ratio and number of samples, rather than the Gaussian modulation profile and thus would affect the symmetrical and asymmetrical protocols similarly. The position of the pessimistic bound for the unmeasured correlation in $p$ in particular does not depend on the post-processing efficiency.

\section{Summary and Conclusions}
We have proposed and investigated the unidimensional continuous-variable quantum key distribution protocol based on the Gaussian modulation of a single quadrature of coherent states of light, in which physicality bounds enable to limit the eavesdropping attacks and assess the security region. The protocol allows simpler technical realization with no need of phase quadrature modulation and full channel estimation at the cost of lower key rate and higher sensitivity to channel excess noise, compared to the symmetrical coherent-state protocol. However, the performance of the protocol is still comparable to that of the symmetrical counterpart and allows for the practical implementation.


\acknowledgements
We thank Eleni Diamanti, Paul Jouguet, and Christoph
Marquardt for the discussion on the experimental interest of
the proposed protocol. We also thank Anthony Leverrier for
drawing our attention to the work of Zhao {\it et al.} The work of
VCU was supported by Project No. P205/10/P32 of the Czech
Science Foundation, by the European Union FP7 Project No.
BRISQ2 under Grant Agreement No. 308803, co-financed by
MSMT\v CR (7E13032), and by the NATO SPS Project No.
984397. Thework of FG was supported by the French Agence
Nationale de la Recherche FReQueNCy project (Project No.
ANR-09-BLAN-0410-02).


\begin{thebibliography}{99}
\bibitem{footnote2}The order in which the limits $V_M\rightarrow\infty$ and $\eta_x\rightarrow0$
  are taken is important and changes the result. In particular, the order taken 
  here assumes $V_M\sqrt\eta\gg1$.
\end{thebibliography}
\end{document}